\documentclass{proc}
\usepackage{amsfonts}
\usepackage{amsmath}
\usepackage{graphicx} 
\begin{document}

\title{Music Data Analysis: A State-of-the-art Survey}
\author{Shubhanshu Gupta  \\
%EndAName
Dhirubhai Ambani Institute of Information and Communication Technology\\
shubhanshu\_gupta@daiict.ac.in}
\maketitle

\begin{abstract}
Music accounts for a significant chunk of interest among various online activities. This is reflected by wide array of alternatives offered in music related web/mobile apps, information portals, featuring millions of artists, songs and events attracting user activity at similar scale. Availability of large scale structured and unstructured data has attracted similar level of attention by data science community. This paper attempts to offer current state-of-the-art in music related analysis. Various approaches involving machine learning, information theory, social network analysis, semantic web and linked open data are represented in the form of taxonomy along with data sources and use cases addressed by the research community.  
\end{abstract}

\section{Introduction}

Music accounts of a significantly large part of online activity today with availability of various online music stores, streaming services, news and podcast services, social networks, and even cloud-based personal music collection. With these developments, we are moving towards an interesting contrasting trend. In days of hard record sales (disc, cassette tapes, records) it was easy to keep track of sales while difficult or impossible to track number of times they were played by the listeners in their music systems. As music increasingly released, distributed, played and discussed online, it has become possible to individually keep track of various aspects by analyzing the data in near real-time. However, with millions of songs, artists, events touched by potentially billions of listeners online; the resulting online activity opens up vast avenues of research for data science community. While interested in exploring research opportunities involving current Big Data technologies, we first attempt to capture current state-of-the-art in music data analysis to clearly identify kinds of datasets, features, analytical techniques that are used by the research community to support various applications and use cases. \par

We structure our survey around use cases and attempt select representative research employing class of techniques. We also structure the comments to reveal the kind of datasets utilized, features extracted, analytical techniques and variations in experiments performed along with the outcome of the effort.

\section{Music Analysis Use Cases and Applications}
\subsection{Prediction and Recognition of Musical Aspects}
Music data analysis is widely used for automated prediction or recognition of various musical aspects like musical style, genre, mood, emotion, onset, melodic sequence, along with predicting the success of a song. Appropriate musical features that are associated with these aspects are identified, extracted, processed and subjected to various analytical techniques for the purpose of prediction. This use case is especially useful for automated tagging of songs, synthesis of new music, and determining potential success that a song might garner. 
\subsubsection{Style}
In our survey, we came across supervised learning and semi-supervised learning algorithms which analyze music. Supervised learning algorithms which include statistical classification methods, contains naive Bayesian, linear classifiers and neural network approaches which can be used to recognize musical style, which is, classifying music being played lyrically, frantically, pointillistically, with syncopation, high, low, quote and blues. The data-set consisted of trumpet performances of various music styles recorded as MIDI from actual performances. Thereby, the data-set had a total of 8 styles, each consisting 25 examples resulting in 1200 five-second training examples. A real-time music style classifier has also been built which employs Naïve Bayesian classifier, linear classifier and Neural networks using 13 low-level features extracted from MIDI data. All the training examples that rest in the dataset were used for classifying improvisational style rather than music feature selection and feature learning \cite{Dannenberg1997} . \par
 
In a separate approach, music style modeling consists of deriving a mathematical model, such as a set of stochastic rules from a set of musical examples. The data-set consisted of several MIDI files, which included polyphonic instrumental and piano music with styles including the early renaissance, baroque music, hard-bop jazz, from eclectic sources. Apparently, some of the regularity can be captured in the composition process by using statistical and information-theoretic tools to analyze musical pieces. And the resulting model can be used to infer and predict music style. The statistics and information-theoretic tools consist of dictionary based methods and selective dictionary based methods. The former, operates by parsing an existing musical text into a lexicon of phrases or patterns, called motifs and then provide a rule that infers which musical object to choose next that would best follow a current past context. It consists of Incremental Parsing algorithm which helps in building a dictionary of distinct motifs. The latter, with the help of Prediction Suffix Trees algorithm builds a restricted dictionary of only those motifs that both, appear a significant number of times throughout the complete source sequence and at the same time, are meaningful for predicting the immediate future \cite{Dubnov2003}. Additional work is being done so as to come up with a more general OpenMusic (it is a Lisp-based open source software and visual programming environment for music composition and analysis) based real time performance systems which can have the prowess to catch the music style, while interacting with several performers and responding at the same time.  \par
\begin{figure}[htbp]
	\centering
		\includegraphics{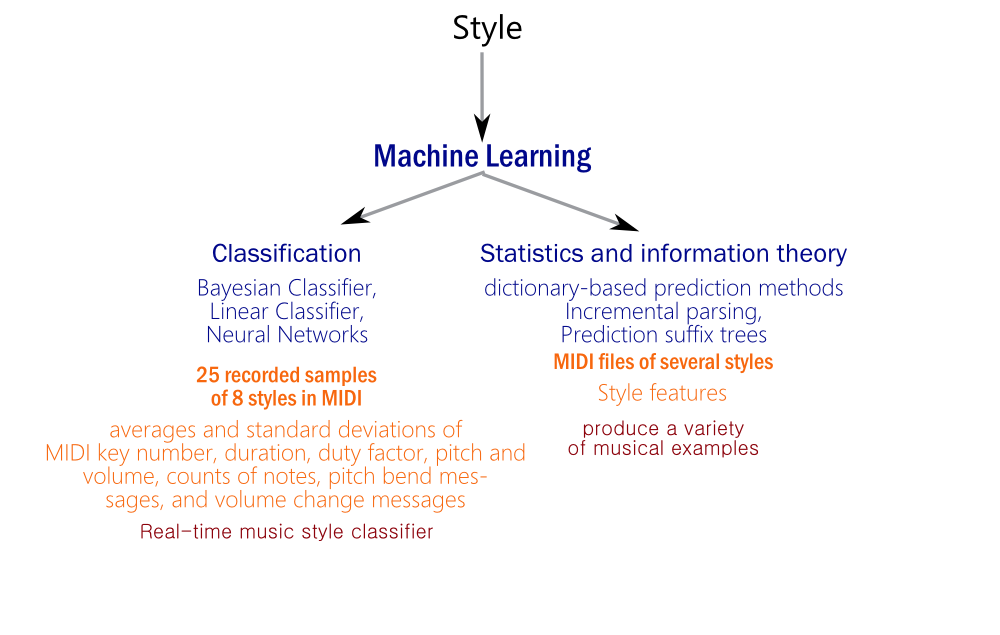}
	\caption{Music Data Analysis approaches applied to style prediction/recognition }
	\label{fig:style}
\end{figure}

\subsubsection{Genre}
Genre detection has been an active area of application in music data analysis. Various state-of-the-art techniques have been applied and reported by the research community. The most frequently referred datasets for carrying out genre classification are GTZAN, ISMIR genre, ISMIR rhythm, and Latin music database containing about 698 to 3227 songs. The most common approach for the application is statistical classification method which consists of KNN, SVM, Random Forest, Naive Bayes and J48 Decision Tree which are used for large scale music genre classification. Significant performance gains have been achieved by beat-aligned vector sequences of the features for large volume of data-sets. In order to capture the temporal domain, using six motley combinations of Echonest features, statistical moments were calculated. Although, there is still a gargantuan room for a large scale evaluation of the remaining features (MFCCs – Mel Frequency Cepstral Coefficients, Chroma, loudness, tempo, key, dance-ability, hotness information etc.) provided by the million song data-set \cite{Schindler2012}. In other attempt, a kNN classifier combined with SMBGT was used for music genre classification of symbolic music. The data-set it used consisted of 100 MIDI songs that span four genre of music namely classical, blues, rock, and pop. There were several loopholes in coming up with a proper music genre classification through SMBGT, similarity measure combined with k-NN classifier, although it was a novel approach. To begin with, instead of just k-NN, a combination of several diverse and independent trained classifiers can be used. Moreover, short segment of musical pieces could have been used for feature extraction. And one of the most prominent space left which can be worked upon is the analysis of polyphonic music instead of just MIDI \cite{Kotsifakos2013}. Music genre recognition at a web scale has been demonstrated using Linked Open Data employing semantic web techniques. Adopting the e-science approach for data and compute intensive jobs, the e-Research infrastructure configured to perform NEMA (Network Environment for Music Analysis) genre classification workflow over Jamendo free music collection dataset that are converted into semantic representations \cite{DeRoure2011}.
\begin{figure}
	\centering
		\includegraphics{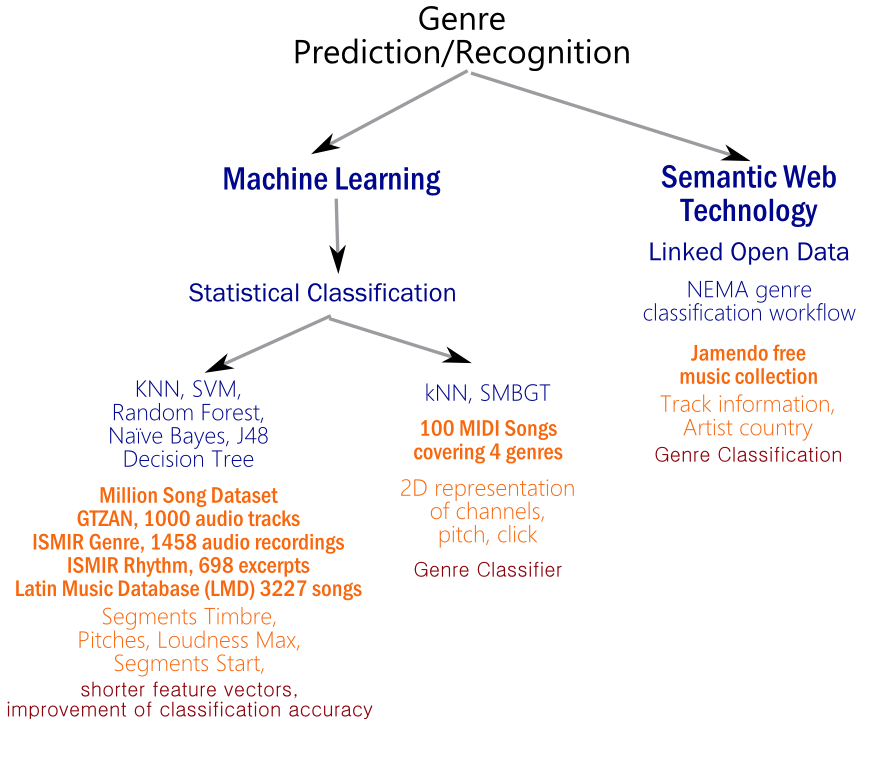}
	\caption{Music data analysis approaches applied to Genre recognition}
	\label{fig:Genre}
\end{figure}

\subsubsection{Mood}
The number of social, personal, and other activities that we are involved in our day-to-day lives is proliferating. And each activity usually arrogates different moods. Hence, usually the music buffs prefer listening to different kind of music that best fits their current mood. And this is where mood classification of songs is gaining utmost importance these days. Some MIR techniques that are used for mood classification of music are the Network Environment for Music Analysis (NEMA) system; the jMIR suite. The MIR solutions that use semantic web technologies incorporate GNAT and GNARQL software tools which use the music ontology and Sonic Visualizer, Annotator tools and their VAMP audio analysis software plug-ins also use the music Ontology.  The methodology of automatic mood classification of songs  also relies on  songs’ data such as lyrics and meta data and the classification is carried out through SVM, Naive Bayes classifiers of supervised learning algorithms and Graph based methods (NB: content based Naive Bayes classifier; GC-Oh: graph-based method by Oh et al; GC-New: graph based method with extension of neighbor function) of semi-supervised algorithms. The data-set consisted of about 6000 songs tagged with mood categories which can consist up to 132 predefined type of moods from a blog site called LiveJournal whereas the whole data of lyrics came from LyricWiki website. It was found that the above used framework and methodology was not sufficient and assertive in mood classification for a real music search engine system. But if proper audio information like artist, sentiment words, more weightage on words in chorus and title parts, on combining with lyrics might fetch better accuracy and results of mood classification \cite{Dang2009}. 

Besides MIR solutions for mood classification of songs, Semantic information retrieval of music is also playing an instrumental role in determining listener’s emotional responses. The framework here, in this case, evolves from low semantic concept level (audio signal) to high semantic concept level (mood).  The inputs extracted from the web are the various social and meta data based information like socio-cultural tags, editorial data, annotations etc. In the web extraction module, since the social media is augmented with ever growing rich context and social meta data information, an SVM based music mood machine learning method helps in the audio feature extraction. And finally, mood annotation is predicted with semantic association via semantic reasoning through TBox, ontology’s terminology and ABox, ontology’s assertional axioms.  The mood-oriented TBox in constructed on the base of Music Ontology terms and refines the specific music-mood which has two main parts: the “Web-based part” which refines high level social meta data information and the “Audio-based part” which refines audio based information. And ABox is constructed with information being extracted from raw audio and web information. The web based information is extracted from all the meta data rich websites such as Last.fm, AllMusic, etc., to get ID3 meta data, tags, annotations, editorial information, comments, etc. The data set consists of about 1804 tracks, which covers about 21 major genres and 56 sub genres and includes 1022 different artists. It was also observed that the accuracy of the mood annotation can improve by a large extent by the embellishment of the other meta data because of the proliferating context based social meta data information, burgeoning from the social media sites \cite{Wang2010}.

Above, wherein mood classification was carried out through the statistical classification methods and Graph based methods of machine learning, it was elicited that combining audio information with lyrical data might yield better results for mood classification and this was addressed in the methodology discussed above in which ontology based methods were used to link the audio information with the available web based information which simply outperformed all the other methods (and hence confirming the speculation about its validity).  
\begin{figure}
	\centering
		\includegraphics{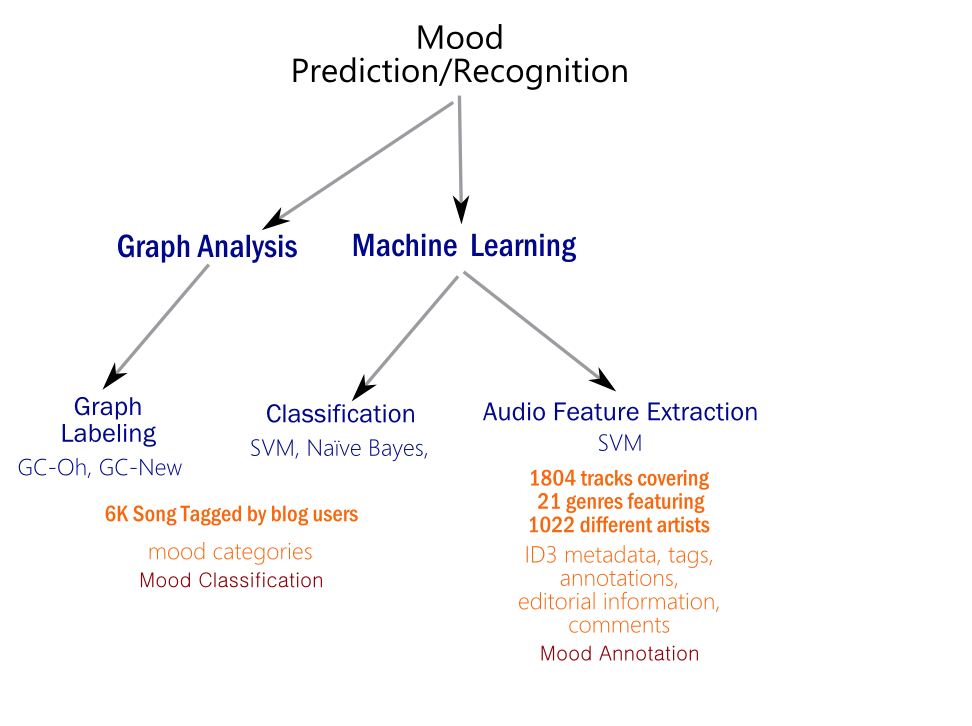}
	\caption{Music data analysis approaches applied to mood analysis}
	\label{fig:mood}
\end{figure}

\subsubsection{Melodic Sequence}
Machine learning is also used for modeling fixed length musical pitch sequences in monophonic melodies. For more comprehensive analysis of sequential structures in music that also includes other musical features and polyphonic structures, musical pitch serves as a starting point. The Restricted Boltzmann Machine algorithm in the artificial neural networks of the supervised learning algorithms of machine learning is used for learning sequences of musical pitch. The data set consisted of 185 J.S.Bach chorale melodies and it was found that, although the neural probabilistic model accomplished modeling musical pitch sequences pretty well but still it is not known that whether predictions can be improved if other musical features, like note durations, intervals etc. are introduced or not. Also, the current model has not been proffered polyphonic music for modeling and analysis and therefore, there is also room for expanding the model for a wider and bigger data set instead of just limiting it to the scope of monophonic melodies \cite{Cherla2013}.
\subsubsection{Onset Detection}
Besides the various kinds and types of music classification, it also is sometimes necessary to get the information about various high level tasks which come under music information retrieval paradigm such as onset detection. Onset detection function helps in figuring out the starting points of various events relevant to music in an audio stream such as beat-tracking, score following and music transcription. Hence there exists a peak-picking algorithm based on artificial neural networks (bidirectional recurrent neural network, here), trained in a supervised manner for common onset detection functions. The data set used for evaluation purpose consisted of 321 audio excerpts covering different types of musical genres, performed on various instruments and having a total length of approximately 102 minutes and 25, 927 annotated onsets \cite{Sebastian2013}. It was found that in comparison to the existing hand-crafted methods such as basic peak selection algorithms based on psychoacoustic theory and heuristics, the spectral flux method (the new neural network based peak picking algorithm) is able to clearly outperform the former for existing onset detection functions.
\subsubsection{Song Hit Prediction}
Machine learning is also used for predicting the success of songs even before they are released in the market, referred to as the “Hit song science”. In this, accurate models are built to predict if a song would be a top 10 dance hit or not, for which a dataset of dance hits was retrieved which included 21,692 instances with five features: song title, artist, position, peak position and date and these instances were retrieved from 3,452 out of 4,120 unique songs in the hit list database. Five machine learning classification techniques, models namely, C4.5 decision tree, logistic regression (which comes under linear classifiers) of statistical classification method and Support Vector machines (SVM), RIPPER ruleset and naive Bayes, are used to build hit song classification. . The results have clearly shown that logistic regression technique fairs the best in comparison to all the other techniques followed by naive Bayes. Although it has been observed and derived that machine learning is an effective measure for educing the top hit songs but the use of music information retrieval systems has not been explored as of now for predicting hit song. There has been some work in determining popularity of a song, based on acoustic, lyric, and human based features, but these factors too have not been able to deliver the results \cite{Herremans2013}.
\subsection{Classification}

\subsubsection{Music Classification}
Many areas of research in MIR involve music classification (genre speech segmentation, emotion chord recognition, playlist generation, audio to symbolic transcription etc.). The fundamental tasks of music classification include musical data collections (called instances) audio recordings, scores, cultural data (e.g. playlists, album reviews, billboard stats, etc.) which also include meta data about the instances like, artist ID, title, composer, performer, genre, date etc. This musical data collection undergoes feature extraction, wherein features represent characteristic information about in-stances and then finally, Machine Learning algorithms (classifiers and learners) learn to associate feature patterns of instances with their classes for music classification.  jMIR, a powerful, flexible and accessible software has been developed to meet the need for standardized MIR research software in order to design, share and apply a wide range of automatic music classification technologies. The jMIR software has been designed to  facilitate  the extraction of meaningful information, available on the web, from the audio recordings, symbolic musical representations and cultural information; it also  uses machine learning techniques  to build classification models automatically; the software also collects profiling statistics and automatically detects meta data errors in musical collection; it also conduct experiments on music collections in both audio and symbolic formats which are a set of large, stylistically diverse and well labeled collections of music; and  it also helps in storing and distributing information in expressive and flexible standardized file formats so as to use that information for automatic music classification . Significant performance gains for music classification was observed when features extracted from multi-modal information like audio recordings, symbolic recordings and cultural data were combined, instead of using features from just one type of data \cite{mckay2010automatic}.

\subsubsection{Similarity}
Today, the amount of music available on various online music stores is continuously increasing in spite of the fact that they already house millions of downloadable songs in their catalog. This leads to a requirement of intelligent music search algorithms to discover and navigate several millions of music for finding their acoustic neighbors. The filter-and–refine method, designed to work with very large databases, is based on FastMap which allows quick music similarity processing. It uses Gaussian timbre models and the Kullback-Leibler divergence as music similarity measure. The data set used is a collection of 2.5 million songs which consists of 30 second snippets of songs. FastMap is a MultiDimensional Scaling (MDS) technique. MDS is a widely used method for visualizing high-dimensional data. The input it uses is the distance matrix of a set of items and the data is mapped to the vectors into an arbitrary-dimensional Euclidean space. Usually higher dimensions yield a better and accurate mapping of the original similarity space \cite{Schnitzer2009}. Since the method for music similarity is designed for Gaussian music timbre features using the symmetric Kullback-Leibler divergence, it was observed that it could be extended and generalized to other distance measures too. \par

Music similarity, which helps to understand why two pieces of music or artists are perceived alike by the listener wherein the listener might be able to state the resemblance between the two songs but not the similarity, has been under a lot of research. Some work has also been done in addressing the measurement of similarity between music artists via features which are basically text-based  and are extracted from web pages. Music similarity does not only help find acoustic neighbors of a particular music but also automated playlist generation, music recommender system, music information systems or intelligent user interfaces to access music collections. Hence, there exists an enormous room for the text-based features extractable from artist-related web pages to be able to contribute in context-based music information (similarity) research.. The dataset for the approach was constructed by querying the search engine for every particular artist and thereby building a collection of web pages which might be in the form of fan pages, biographies, album reviews, track lists, etc. No matter how many web pages are retrieved for an artist, the whole collection is considered as one large, virtual document describing the artist for whom the web pages have been extracted. And therefore, web-based music similarity estimation revolves around constructing text-based feature vectors for IR purposes, for example- term frequency, inverse document frequency, virtual document modeling, normalization with respect to page length, similarity function. The term frequency of a term in a document estimates the importance the term carries for the document (related to artist). The inverse document frequency estimates the overall importance of the term in the whole corpus. Virtual document modeling relates to the way individual documents are aggregated, retrieved for the same artist. The different similarity functions come up with the estimation of the proximity between the term vectors of two documents or artists. But the interdependency between these leads to a problematic situation wherein it becomes difficult to choose which variant (e.g., variant of term frequency, variant of similarity measure) would produce an overall winning combination. But each variant focuses on the task of text-based similarity estimation of music which is a specific, important task of music information research \cite{Schedl2011}. It was also clinched that the above methodology also possesses latency for the development of personalized music retrieval system. Above, only text-based representation of music data derived from artist web pages has been mentioned but it is also possible to consider in the data which burgeons from user-generated content like instant messages or posts and updates made on social networking websites. This will help in devising better music similarity with the impetus of social factors. The inclusion of this new dataset suggested above might also help in improvising music playlist generation systems. \par

Metric Learning to Rank (MLR) is an extension of the Structural Support Vector Machines (SVM) approach which can be applied for learning a Mahalanobis distance that, according to the relative similarity ratings by users, casts an apprehended or stated music similarity based on the MagnaTagATune dataset for acoustic recordings of music, and can be applied in music exploration or recommendation systems. The MagnaTagATune data set comes from TagATune, a web-based game which collect tags associated with certain songs in a human-computation manner. MagnaTagATune dataset consists of features and tagging information of 25863 29-second audio clips generated from 5405 source MP3s \cite{Stober},  \cite{Detyniecki2013} . It was also observed and proved that the methodology, stated above, need not be only restricted to the proposition that models distance as a weighted linear combination of facets rather it can also be extended to follow the approach that incorporates Mahalanobis distance. \par

Music similarity techniques also help in music retrieval and recommendation. By developing methods to identify and extract relevant entities (e.g., artists, full names, band line-up, album and track titles, related artists, etc.) and relationship between these, a better and improved multi-faceted similarity measures can be strived for. And this leads to a possible solution for determining members of a music band, i.e., which persons a music band consists (or consisted) of by analyzing texts from the web and taking the fact for granted that, any person that has been a member of a band at any point is considered to be a band member. Band member detection is a case of named entity recognition which comprises of the identification of proper names as well as the classification of these names achieved through rule-based approach or supervised learning approach. There are two rule-based approaches: Hearst Pattern Approach which automatically extracts the line-up of a music band, whereby, line-up information includes the member information and the corresponding roles that they are playing, for example, the instrument that they are playing. Another rule based approach uses the GATE, an open source framework (General Architecture for Text Engineering), automatically identifies artist names, to extract band-membership relations, and to extract released albums/media for artists. After the rule-based approach, the supervised approach for band member detection  can be listed as: Named Entity Recognition in GATE; followed by extracting band members by supervised learning algorithms such as hidden-markov-models, decision trees, or support vector machines (SVM) wherein the SVM is chosen as a classifier. So as to extract the band members by supervised learning algorithms with SVM as classifier, the data set is constructed, first, on querying 51 rock and metal band members on Google and thereby getting a total of 5,028 web pages. Secondly, the data set consists of band biographies fetched from band-membership information of 34,238 bands with the help of Echonest API leading to a total of 38,753 biographies. . Now, construction of features takes place wherein two distinct SVM classifiers are trained so as to detect person entities to be marked as band members.  Then entity extraction is carried out (to detect band members and assign a confidence score) followed by entity consolidation and member prediction in which a list of potential band members is obtained from the named entity extraction step for each processed text. On the metal page set (web pages derived after querying the search engine), the advanced rule-based approach performs better than the supervised learning approach whereas the case is opposite in case of the biography set where the supervised learning approach performs better \cite{Knees}. With the methodology discussed above for finding out the band member, it is also possible to generate really coveted meta-information on music. Every biography consists of information on the band members, the composers, musicians, vocalists, guitarists (if any), etc. If all of this semantic information is extracted and properly annotated, not only it will provide unprecedented solutions in music retrieval and recommendation but it will also be able to give proper credits (and hence royalty) to every artist who contributed in any way for the making of a song. \par    

\begin{figure}
	\centering
		\includegraphics[width=0.50\textwidth]{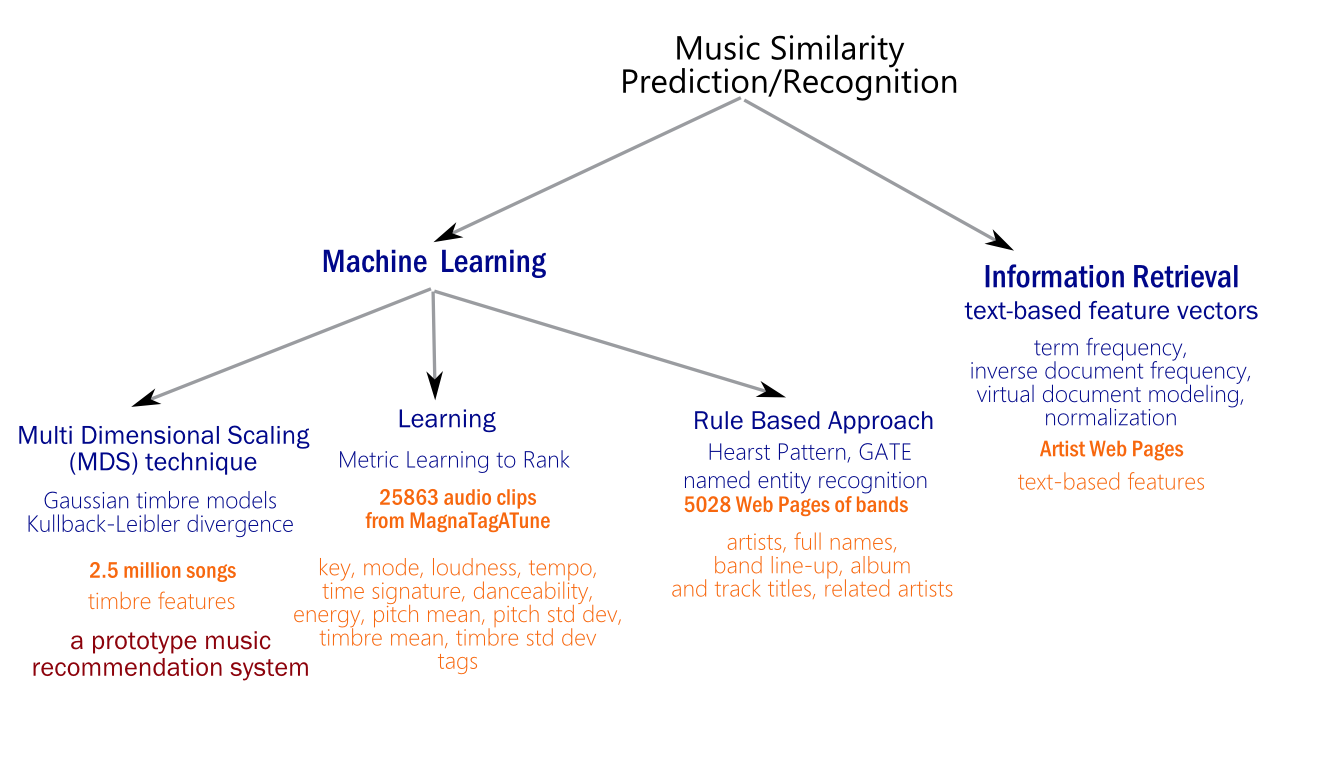}
	\caption{Music data analysis applied to similarity }
	\label{fig:Music Similarity}
\end{figure}

\subsubsection{Emotion}
Besides fields like music genre identification, mood detection, style recognition, and even music similarity recognition, music emotion recognition is seeing compounding growth in research interest because music enjoys a prominent status in human lives because of its ability to elicit emotions which are subjected to our mood and changes in physical condition and actions. It is possible to do music emotion recognition by a method based on melodic features extracted from polyphonic music excerpts through machine learning algorithms. The dataset used was a set of 903 audio excerpts, each of 30-seconds organized in 5 relatively balanced clusters of 170, 164, 215, 191, 163 excerpts respectively. Several supervised learning algorithms namely Support Vector Machines (SMO, LibSVM), K-Nearest Neighbors, C4.5, Bayes Network, Naïve Bayes, and Simple Logistic of machine learning were applied and ran on Weka, a data mining and machine learning platform with best results being achieved using SVM classifiers \cite{Rocha2013}. The methodology discussed above was applied on Melodic Audio features which can be of three types namely pitch and duration, vibrato and contour typology. And it was found that if, along with Melodic Audio features, Standard Audio features (it includes spectral shape features like centroid, spread, bandwidth, skewness, etc. which come under low level descriptors and tempo, tonality, key etc. which come under high level descriptors) are also incorporated for music emotion classification, not only performance might increase but the accuracy of the classification might also increase.  

The initial works in music emotion recognition used an audio based approach that demonstrated music being associated with discrete emotion categories. Features such as timbral, rhythmic, and pitch trained in Support Vector Machines (SVM) leads to large variations in the accuracy of estimating the different categories. Black Propagation Neural Network (BPNN) recognizes the extent to which the music pieces belong to four emotional categories namely, happiness, sadness, anger and fear. Two datasets, the CAL500 and the other consisting of approximately 21000 clips from Magnatune modeled using statistical distributions of spectral, timbral and beat features using Multi-Label k-Nearest Neighbors (MLkNN), Calibrated Label Ranking (CLR), Backpropagation for Multi-Label Learning (BPMLL), Hierarchy of Multi-Label Classifiers (HOMER), Instance Based Logistic Regression (IBLR), and Binary Relevance kNN (BRkNN) models. It was found that CLR classifier using a Support Vector Ma-chine (SVM) outperformed all other approaches besides performing competitively with Decision Trees, BPMLL and MLkNN \cite{Barthet2012}. It was found that in order to improve the efficiency of the music emotion recognition, rather than just low-level descriptors (which revolve around tempo-related aspects of a song), mid or high level descriptors need to be incorporated which carry semantic or syntactic meaning like genre and culture, moods and instruments, or rhythm and tempo. Also, most of the current approaches employed in the music emotion recognition do not account the gravity of the relationships that exists between features and emotion components and hence lose out on unsullied music emotion recognition. Moreover, there also lies great scope of adopting semantic web ontology in this field which has not been delved into as of now. It has also been proven that if a multi-model music emotion recognition model is built capitalizing on audio content and semantic association reasoning, it is bound to give promising results in performance. Hence, there lie immense possibilities which can rummage better yields in music emotion recognition.

There also exists ways to maximize the performance of a music emotion recognition system based on regression approach of machine learning. The data set consisted of 50 ratings per clip for 288 clips where clip is an excerpt of a track. The various regression algorithms that approach music emotion recognition differently are: Linear Regression (LR) – assumes linear relationship between input and output variables and minimizes the least square error; Regression Tree (RT) – here each leaf node is a numeric value rather than a class label; Locally Weighted Regression (LWR-SLR) – constructs a one factor linear model on-the-fly based on nearby training points when presented with a test sample; Model Tree (M5P) - here each leaf node is a linear mod-el rather than a numeric value and it contains a number of parameters that must be optimized during training with a parameter search; Support Vector Regression (SVR-RBF) – implemented in LIBSVM using the Radial Basis Function; Support Vector Regression with No Parameter Search (SVR-RBF-NP) -  here the parameter values are hardcoded to sensible defaults otherwise the rest is same as SVR-RBF. Using all standardized features and a coarse-grid search for the best parameters, the regression SVR with RBF kernel performs the best \cite{Huq2010}. It was found that the above methodology achieved zenith in music emotion recognition. But in order to scale up the desired output, there need to be some efficient technique for gathering colossal data sets properly annotated with emotion labels. Then higher-level music features are required which can be ascertained by human music cognition. Also, certain models of temporal evolution of music can definitely play certain role in advancement of music emotion recognition models along with the development of personalized systems that can predict the various emotional characteristics and responses of people of culturally diverse backgrounds, tastes and various other characteristics.
\begin{figure}
	\centering
		\includegraphics{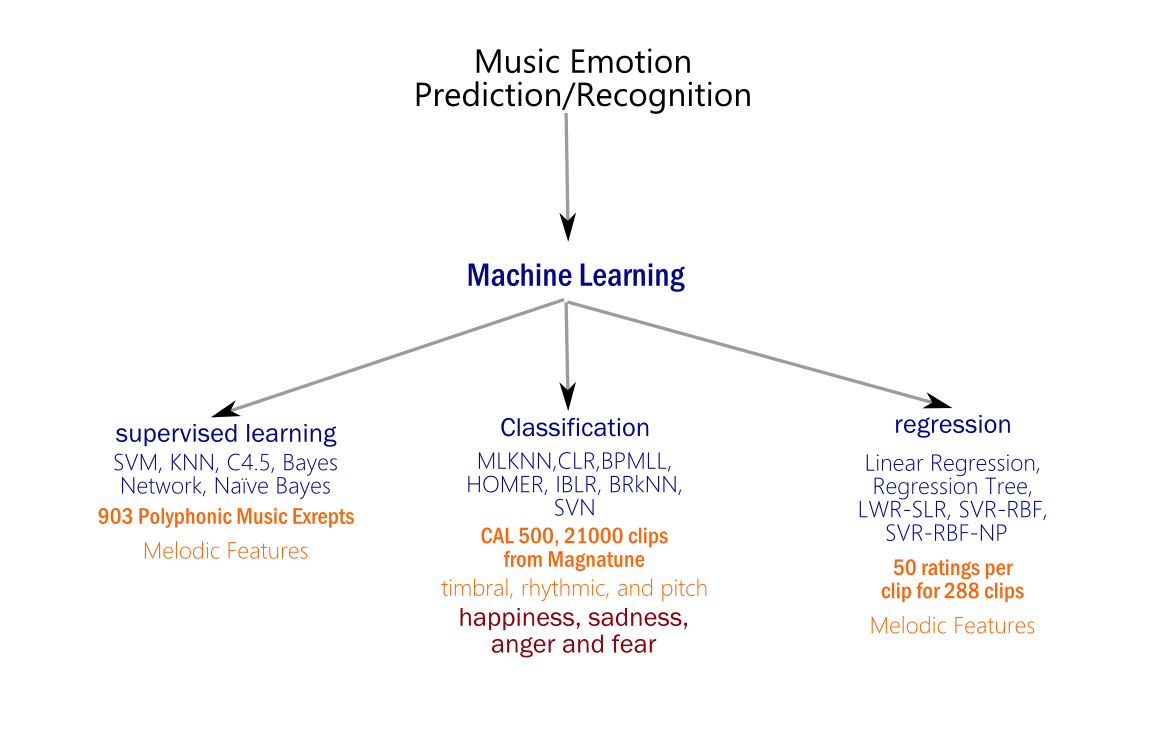}
	\caption{Music data analysis applied for music emotion recognition}
	\label{fig:Music Emotion}
\end{figure}

\subsection{Audio Analysis}
On the similar grounds as that of jMIR, for audio analysis and audio based Music Information Retrieval, there exists an open-source, cross platform C++ library, Essentia 2.0 under the Affero GPL License. The library houses an extensive collection of reusable algorithms for implementing audio input/output functionality, standard digital signal processing blocks, statistical characterization of data, and a large set of spectral, temporal, tonal and high-level musical descriptors. Essentia provides algo-rithms for: basic processing of audio streams so as to achieve audio input/output filtering; for computation of low-level spectral descriptors; computation of time-domain descriptors; computation of tonal descriptors; computation of rhythm descriptors; computation of SFX descriptors; and in addition to all the above low level descriptors, Essentia also provides algorithms for various mid- and high-level descriptors. Essentia has been used for various research activities, with its major contribution in musical classification, mood classification, and semantic auto-tagging, music similarity and recommendation, visualization and interaction with music, sound indexing, detection of musical instruments in polyphonies, cover detection, instrument solo detection and acoustic analysis of stimuli for neuroimaging studies \cite{Bogdanov2013}. Essentia library might witness an update for real time applications and addition of new semantic categories in the set of high-level classifier based descriptors. 
\subsection{Recommendation}
\subsubsection{Music Recommendation}
SoCo, is a context aware recommender system, that incorporates thoroughly processed social network information, recommends music based on the application of random decision trees algorithm of statistical classification methods of machine learning, and takes into account various contextual information like the characteristics of a user under static context which includes user’s age, gender, membership, role etc. or an item’s category, cost, physical properties, etc. and dynamic context which is associated with a ratings’ spontaneous information which might include a user’s mood or his/her location while rating an item (music), and social factors for making personalized and accurate music recommendations since it brings a new perspective in recommendation because social factors bring a whole new lot of information about a user’s preference for an item (music) which can be implicative from the user’s social circle of friends and followers who are counted upon to share similar taste profiles \cite{Federale}. 

%The data set that was chosen for building and testing the methodology of SoCo, was Douban – the largest Chinese social platform for reviewing, rating and recommending books, movies and music. Since, music department was of interest, hence 1,387,216 numbers of ratings, 23,822 users, 185,574 numbers of musical items were taken into consideration from the dataset. The experimentation of the methodology employed by SoCo was proved to be unprecedented and it outperformed the standalone models based on context-aware and social recommendations. It would be interesting to see if SoCo is able to make some impact on any real life project like making efficient web content recommendation, for web content credibility evaluation systems. 

dbrec-a music recommendation system based on Linked Data, has been built on top of DBpedia (it was chosen for two main reasons - the availability of data of more than 39,000 artists, and secondly, availability of pictures and description of artists which is useful for building system’s user interface), which proffers recommendation for more than 39000 bands and solo artists. It uses LDSD algorithm – Linked Data Semantic Distance – as a basis for its recommendation engine. The system is built, on the roadmap which necessitated the need of: identifying the relevant subset from DBpedia; followed by reducing the dataset so as to optimize the query process ; then comes computing the distances using the LDSD algorithm and representing them using its ontology; and to delineate the recommendations, building of a user-interface comes as a last step for browsing recommendations \cite{Passant2007}. 

It’s also possible to let people find and recommend music and its content based on what they are consuming or producing by leveraging social music data to the semantic web. Rather than going with conventional music recommendation practices like collaborative filtering (recommending music to a user based on the stated tastes of other related users), content-based, and recommendation by modeling musical audio similarity, relationships between various types of data (social networks, published content, tags, artist information, etc. ) which are modeled in RDF from the social music websites. It is achieved by: interlinking FOAF (Friend Of A Friend) and linked data with various social networks so as to provide a complete distributed and open social graph, that can be queried and processed; SIOC – Semantically-Interlinked Online Communities – is a shared semantics in order to represent user-generated data coming from various places in a common way, by offering a model to represent activities of online communities and their contributions; MOAT framework that allows people to tag their content with URIs, rather than simple keywords, and once people have tagged their data, relationships between those URIs can be used to suggest related data. For example- when browsing a blog post about The Clash, the above mentioned recommender system would recommend to browse a picture tagged with the URI of Joe Strummer on Flickr, because both the blog post and the picture have a relationship defined in DBpedia. In this way, FOAF ontology is reused by SIOC and MO (Music Ontology) and the linking between SIOC data to LOD (Linked Open Data) URIs is allowed by MOAT \cite{Passant}. 
%But still, there exists a large latency in the process of browsing the graph of linked data to provide good recommendations which can be improvised manifold times. Using a user’s profile data which might include attention and taste data, political interests, interest in other arts etc., can prove to be a nascent step in improvising the traversal on the graph of linked data. Moreover, the whole music recommendation approach discussed above was based on data available from DBpedia, whereas other datasets can be incorporated for experimentation as well.  

One of the challenging aspects of music recommendation is to implement situation-aware personalized music recommendation service which takes both the user situation as well as user preference into consideration. This requires multidisciplinary efforts which includes human mood and emotion recognition from low level features (like beat, pitch, rhythm and tempo) extraction and analysis. Hence, a new scheme, Context-based Music Recommendation (COMUS) ontology, was devised for situation aware/user-adaptive music recommendation service in semantic web environment. 
%COMUS, Context-based Music Recommendation, can model a user’s musical preferences; context and his/her desired emotions and preferences. It basically defines an upper Music Ontology (it stores ABox knowledge as OWL instances) which captures concepts on the general properties of music such as title, artists and genre. This ontology is defined in Web Ontology Language (OWL). OWL provides three increasingly expressive sublanguages: OWL Lite, OWL Description Logic (DL), and OWL Full. A DL-based reasoning engine is able to answer semantic queries with regards to various types of mood and situation (this ontology have TBox knowledge as concepts) and for recommendation purpose, SPARQL query language is used in conjunction with the music extended OWL ontology to find out appropriate music. Hence, the overall module looks into three main components: COMUS along with ontology, recommendation module based on score, and mood recognizer. 
COMUS provide various query interfaces to the user: Query By Situation (QBS), Query By Detailed Situation (QBDS), and Query By Mood (QBM). The Jena SPARQL engine is used for fetching and recommending purposes. It is used for expressing and processing necessary queries to the ontology \cite{Rho2011}. The dataset used for the above methodology is counted by the number of RDF triples, each representing a subject, verb and an object. Thereby, the COMUS ontology is a huge collection of 826 OWL classes and instances parsed into 3645 RDF triples. The methodology covered the aspect of presenting and building ontology based on context modeling and reasoning for the purpose of music recommendation by modeling musical domain and capturing low-level musical features and factors which represent various music moods and situations like time and location which can influence craving for different types of music. 
\begin{figure*}
	\centering
		\includegraphics{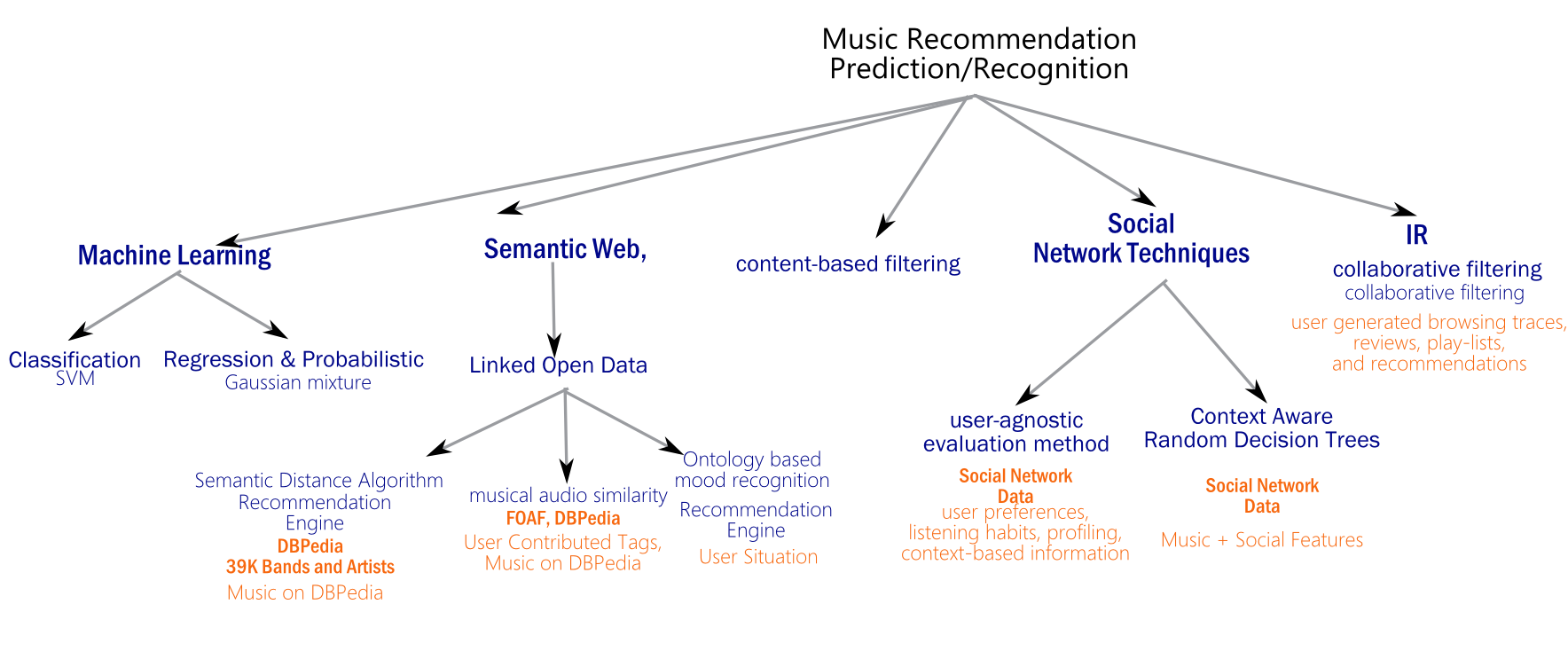}
	\caption{Music data analysis applied to recommendation}
	\label{fig:Recommendation}
\end{figure*}

Music Information Retrieval has also been a consistent approach for music search and recommendations including the search for items related to a specific query song or their set. There are various online communities which provide a huge amount of user generated browsing traces, reviews, play-lists, and recommendations which can be analyzed through collaborative filtering methods so as to generate relationship between artists, songs and genres. These relationships then in turn result in recommending music to users based on their music activity  \cite{Brandenburg2009}.

There also exist some unprecedented methods of recommending music. A user-agnostic evaluation method (or network based evaluation applied to artists and large scale user similarity graphs) which is based on the analysis of the item (or user) similarity network, and the item popularity. 
%It basically, measures the novelty component of a recommendation algorithm, uses the complex network analysis so as to analyze the similarity graph, models the item popularity curve and then combines both the complex network and the item popularity analysis so as to be able to come up with the characteristics of the recommendation algorithm without any user intervention. Then there is a user-centric evaluation method which is based on the immediate feedback of the recommendations supplied. In this method, there is a provision which allows users to provide immediate feedback to the evaluation systems, so the system can react accordingly and thereby measuring the novelty factor and the relevance of a recommendation algorithm. This method complements the previous method and is used for evaluating three different music recommendation approaches (social-based, content-based, and a hybrid approach which uses expert human knowledge). 
There is a system prototype, named FOAFing the music, which provides music recommendations based on the user preferences, listening habits, profiling, context-based information (which is extracted from music related RSS feeds), and content-based descriptions (which is automatically extracted from the audio itself). Then there is a music search engine, named Searchsounds, providing keyword based search, as well as the exploration of similar songs using audio similarity and thereby allowing users to discover music, even unknown  to them  \cite{Herrada2008}. 
%The various approaches presented above have a serious limitation of being derived from a dump of the dataset (item similarity network) instead of, from a dynamic dataset. It is because, user’s taste fluctuate and even change permanently over the course of time and even in the same geographical area and so does the similarity between the items analysed changes. Hence, it becomes important to derive results and build systems with a dynamic model in hand. Also, the listeners too can be divided into groups depending upon their inclination and frequency to listen to music. Recommender systems designed for casual listeners would seldom digress from recommending the right type of music but they might in case of music enthusiasts or savants. 

Context-aware music recommendation retrieves and suggests music depending upon the user’s actual situation, for ex-his/her emotional state (can be influenced by age, gender, personality traits, socio-economic, cultural background, etc.) which varies time to time and is pretty complex to be understood by a machine, influences the user’s perception of music. Its importance in music recommendation has led to more refined research for better results. There are some approaches which improve the quality of recommendation: Collaborative filtering (CF) relies on user-generated con-tent (ratings or implicit feedback) - items are recommended to a user if they were liked by similar users. The dataset used in this case was derived from Last.fm social network that delineates a weighted social graph among users, the tracks they play, the tags they annotate the tracks with. Then there is Content-based approach that relies on traditional music information retrieval techniques like acoustic fingerprint or genre detection. 
%The method stores information describing the items and then the items that are known or anticipated to be liked by the user are retrieved. The key step of this approach is learning the user model based on their preferences. Nearest Neighbor and the Relevance Feedback approaches are learning algorithms used for the above purpose. Then there is a hybrid approach (to take advantage and avoid the shortcomings of collaborative and content-based approaches, .i.e., new items/new user problem, and the problem of modeling user’s preferences). 
The hybrid approach incorporates the following techniques: the scores that come up as a result of various techniques, they are combined to produce a single recommendation; system switches its judgment of recommendation based on certain criteria, ex- dataset properties, quality of produced recommendations; all the different techniques that produce recommendations are mixed and presented together; item features like ratings and content features from different recommendation techniques are thrown together into a single recommendation algorithm; one recommendation technique refines the output of another technique, for ex- CF can be used to produce a ranking of the items and then the content-based filtering can be applied to break the ties; one recommendation’s output act as a input for another, for ex- CF might be used to find items relevant for the target user and this information is used in the content based approach; and the model learned by one recommender acts as an input for the other – this approach uses one system to produce a model as input for the second system. Since, it is not clear which of the two (CF or content-based approach) has a bigger impact on quality recommendation, it’s best to mix the two techniques in hybrid approach for music recommendation  \cite{Kaminskas2012}. 
%Semantic gap, the anachronistic obstacle in building context-aware music recommendation systems, that is, knowing how people perceive music depending upon various factors such as social, geographical to name a few. In order to bridge the semantic gap for an efficient music recommender system, more work still needs to be done in cognitive psychology, i.e., studying human perception for music; affective computing which includes music emotion recognition; social computing which is nothing but exploiting user-generated content for music information research. 

Another aspect of music recommendation, using heavy machine learning, is the auto-matic prediction of tags to music and audio for music recommendation. Applying tags (a user-generated keyword) in music can be understood as say, listener likes rock music with female voices. The dataset used for this comes from various sources: Social tags – are the ones applied by humans on artists, albums or a song. It was gathered from sources like Last.fm which contains more than 960,000 free-text tags and millions of annotated songs; Games – there are various tagging games developed in order to gather clean data of tags. The Magnatagatune dataset contains tags applied to about 20,000 songs which is the largest game data made available; Web documents – documents available on the internet can also be used to describe audio but it contains a lot of noise. 
%Other methods include querying search engines and mining music websites. The automatic tagging algorithm requires a model that could link tags to audio features. This model can be seen as a product of machine learning (ML) algorithm which then can be applied to automatic tagging. The ML methods used here can be divided into two categories: classification methods and regression and probabilistic methods. Classification methods include: Support Vector Machines (SVM) is applied to automatic tagging. It provides supervision for training at the level of collection of clips (ex- tracks, albums, or artists) instead of providing at the level of the clips; Boosting – is a meta-algorithm in the sense that it works on the top of other learning algorithms. It finds the best possible classifier and adds it to a strong classifier so that classifying the hard ones is more rewarding on reweighting the examples. Regression and Probabilistic methods include: Gaussian mixture - it models a distribution using Gaussians. The advantage of this model is that it gives a probability, i.e., a probability that a tag X has been applied to an audio frame. It also is more powerful than a normal algorithm (in terms of representation capacity) because it estimates the likelihood of a data point. One disadvantage of this method being, that it is extremely demanding in terms of computational resources. In order to overcome this, Hierarchical Gaussian Mixture Model (HGMM) is used which trains Gaussian mixture on a subset of audio frames. 
KNN (K-Nearest Neighbors) is also one of the simplest and effective ML technique used for automatic tagging algorithm. Neural Networks of ML handles multi-label classification and regression cases and thereby is able to capture highly complex relations between audio and tags. HGMM, SVM, and Boosting are one of the three best performing algorithms on automatic tagging \cite{bertin-mahieux09}. 
%Automatic tagging of audio through machine learning have given unprecedented solutions in this field and it can be perceived as a tread towards understanding of music by machines but this field of research needs: the use of large and scalable and really fast algorithms so as to be categorical enough to handle large amount of data in the form of meaningful information; along with proper handling of sparse coding, esp. in time domain because of limitations in the frame based audio features like tradeoffs for frequency or time accuracy. Audio fingerprinting is one of the most notable application of sparse coding.  

\subsubsection{Playlist Recommendation}
Similar to SoCo, there exists another system called MyMusic that exploits social me-dia sources for generating personalized music playlists. It is based on the information extracted from social networks, like Facebook and Last.fm for carrying out the personalization tasks of defining a model of user interest based on a user’s information related to music preferences on social networks. The social media based playlist is enriched with new artists related to those the user already likes. And specifically, two enrichment techniques are used: in the first one, the knowledge stored on DBpedia is leveraged, whereas in the second one, it’s based on the content-based similarity be-tween descriptions of artists. Thereafter the final playlist is ranked accordingly and presented to the user for listen to the songs and for feedback \cite{Musto2012},

It is also possible to contextualize playlist, a set of songs, as a recommendation engine with the help of a novel multi-model similarity measures integrating content-based similarity with artist relational social graphs. In an attempt to evaluate the application (driving a user-steerable radio station by using complex similarity and community segmentation), playlists are compared on a novel low-dimensional song level feature using social tag descriptors which greatly improvise the understanding and construction of playlists for music recommendation. The dataset used for the above mentioned techniques is gathered from radio station logs. Data from yes.com consisted of 885810 number of song entries, 2543 number of songs which had no tags attached to them, 70190 total numbers of playlists, 55 minutes of average runtime of these playlists and 12.62 mean numbers of songs per playlists. Data from Rock stations consisted of 105952 number of song entries, 865 number of songs which had no tags attached to them, 9414 total numbers of playlists, 53 minutes of average runtime of these playlists and 11.25 mean numbers of songs per playlists. Data from Jazz stations consisted of 36593 number of song entries, 1092 number of songs which had no tags attached to them, 3787 total numbers of playlists, 55 minutes of average runtime of these playlists and 9.66 mean numbers of songs per playlists. Data from Radio Paradise consisted of 195691 numbers of song entries, 2246 numbers of songs which had no tags attached to them, 45284 total numbers of playlists, 16 minutes of average runtime of these playlists and 4.32 mean numbers of songs per playlists \cite{Fields2011a}. 

%Inspite of such novel methods for musc playlist generation, there are still many potholes which need to be filled otherwise this field would continue trembling. There is a requirement to have full knowledge and scale of an artistic network, along with increased reliance on computationally complex audio-derived features and a proper solution which can cater to the issues circling around the evaluation of playlists.

\section{Discussion}
In this investigation, we attempted to identify various aspects of music data analysis as addressed by the research community. Datasets used for various analysis reported here were identified and stated clearly. In most cases, datasets consists of relatively small number of audio files in the form of MIDI sequences, user generated tags, accompanying web pages, or user context. While these types of musical datasets are certainly critical part of music domain, many other aspects remain untouched by such research efforts. Some of these include: music credits data; licensing and rights data; digital supply chain data; music sales and distribution data; cataloging, classification and archival related data; music organization data; music-related standards data; live events related data; and studio recordings related data among many others generated throughout the lifecycle of a music professional. These datasets can be maintained by various organizations at multiple levels of details, accuracy and update frequency with potential overlaps. As majority of these data sources are updated constantly at varying frequency, the task of integration and management of datasets itself will require application of appropriate Big Data technologies currently available.  
Next important factor is the features extracted from datasets that can be subjected to analysis selected as per the application or use case requirements. In this survey we identified various high and low level features typically being used by the community. However, various additional musical features can be identified from extended lists of datasets identified earlier. Applying known analytical techniques over these novel features will open up opportunities for novel applications and use cases. 

\section{Conclusions}
In this paper we attempted to offer a state-of-the-art survey of research efforts involving music data analysis. Our objective is to investigate a report various analytical approaches adopted by the research community focusing on unique musical features. This resulted in depiction of a technology landscape of analytical techniques including machine learning, semantic web, social network analysis, information retrieval statistics and information theory. Our analysis also identified opportunities for further exploration keeping in mind new possibilities offered by the recent developments in Big Data discipline. 

%\begin{thebibliography}{9}
%
%\end{thebibliography}

\bigskip

%\appendix
%
%\section{The First Appendix}
%
%The appendix fragment is used only once. Subsequent appendices can be
%created using the Section Section/Body Tag.

\end{document}